\documentclass{article}

\usepackage{cite}
\usepackage{PRIMEarxiv}
\usepackage[utf8]{inputenc} 
\usepackage[T1]{fontenc}    
\usepackage{hyperref}       
\usepackage{url}            
\usepackage{booktabs}       
\usepackage{amsfonts}       
\usepackage{nicefrac}       
\usepackage{microtype}      
\usepackage{lipsum}
\usepackage{fancyhdr}       
\usepackage{graphicx}       
\usepackage{listings}
\usepackage{xcolor}
\usepackage{threeparttable}
\graphicspath{{media/}}     
\usepackage{booktabs}
\usepackage{multirow}
\usepackage{hhline}
\usepackage{subfig}
\usepackage{adjustbox}
\usepackage{longtable}

\lstdefinestyle{yaml}{
     basicstyle=\color{blue}\footnotesize,
     rulecolor=\color{black},
     string=[s]{'}{'},
     stringstyle=\color{blue},
     comment=[l]{:},
     commentstyle=\color{black},
     morecomment=[l]{-}
 }
\title{Remote Bio-Sensing : Open Source Benchmark Framework for Fair Evaluation of rPPG
\thanks{\textit{\underline{Citation}}: 
\textbf{Authors. Title. Pages.... DOI:000000/11111.}} 
}

\author{
  Dae-Yeol Kim, Eunsu Goh, KwangKee Lee \\
  Innopia Tech \\
  Gyeonggi-do, Korea\\
  \texttt{\{wagon0004, dmstn96, kklee\}@innopiatech.com} \\
   \And
  JongEui Chae, JongHyeon Mun, Junyeong Na, Chae-bong Sohn  \\
  Electronics \& Communications Engineering, Kwangwoon University \\
  Seoul, Korea\\
  \texttt{\{paperc, mjh110311, najubae, cbsohn\}@kw.ac.kr} \\
  \And
  Do-Yup Kim\\
  Department of Information and Communication AI Engineering, Kyungnam University\\
  Gyeongsangnam-do, Korea\\
  \texttt{doyup09@kyungnam.ac.kr}\\
}

\begin{document}
\maketitle

\begin{abstract}
rPPG (Remote photoplethysmography) is a technology that measures and analyzes BVP (Blood Volume Pulse) by using the light absorption characteristics of hemoglobin captured through a camera. Analyzing the measured BVP can derive various physiological signals such as heart rate, stress level, and blood pressure, which can be applied to various applications such as telemedicine, remote patient monitoring, and early prediction of cardiovascular disease. rPPG is rapidly evolving and attracting great attention from both academia and industry by providing great usability and convenience as it can measure biosignals using a camera-equipped device without medical or wearable devices.

Despite extensive efforts and advances in this field, serious challenges remain, including issues related to skin color, camera characteristics, ambient lighting, and other sources of noise and artifacts, which degrade accuracy performance. We argue that fair and evaluable benchmarking is urgently required to overcome these challenges and make meaningful progress from both academic and commercial perspectives. In most existing work, models are trained, tested, and validated only on limited datasets. Even worse, some studies lack available code or reproducibility, making it difficult to fairly evaluate and compare performance. Therefore, the purpose of this study is to provide a benchmarking framework to evaluate various rPPG techniques across a wide range of datasets for fair evaluation and comparison, including both conventional non-deep neural network (non-DNN) and deep neural network (DNN) methods.\\
GitHub URL: \url{https://github.com/remotebiosensing/rppg}
\end{abstract}


\section{Introduction}
Remote photoplethysmography (rPPG), a technology that employs cameras to capture facial images and measure blood volume pulse (BVP) based on optical principles, has gained significant attention in recent years. The analysis of BVP allows rPPG to assess various vital signs, including heart rate (HR), respiration, and blood pressure. The principal advantage of rPPG lies in its high usability and convenience, which are derived from its minimalistic requirements. It merely necessitates a device equipped with a camera to measure these vital signs.

There has been a surge of interest in rPPG research, resulting in a multitude of published papers and datasets. In response to this trend, the remote physiological signal sensing (REPSS) challenge was conducted at both ICCV \cite{repss1st} and CVPM \cite{repss2nd}, while the vision for vitals (V4V) challenge occurred at ICCV \cite{V4V1st}. More recently, clinical validation and trials for rPPG are being widely undertaken by various institutions \cite{clinical,hospitaltrial}. As reported in \cite{cvd}, rPPG demonstrates the capability to accurately measure HR in subjects with cardiovascular disease (CVD).

The exploration of rPPG can be segmented into three principal perspectives: dataset, preprocessing, and model training. From a dataset perspective, rPPG algorithms should account for a variety of characteristics such as subject movement \cite{motionICE,motionRR,BCGMotion,motionRobust,realworldmotion}, region of interest (RoI) \cite{region+deepphys, assesmentROI, po2018roi, kwon2015roi, lempe2013roi,niu2017roi,zhao2022roi, van2014roi,calvo2015roi}, color space \cite{jaiswal2023color}, light intensity \cite{lowlight,colorillumination,lowlightimageENhancement}, as well as gender and skin color \cite{genderskin, skintonestyle}. While diverse datasets are emerging, only a few comprehensively encompass these factors.

With respect to preprocessing, facial cropping methods can be categorized into (i) cropping based on the size of the face (typically a certain multiple of the face's size), and (ii) tracking facial movement. After facial cropping, different processing methods are applicable, such as (i) continuous method, using the cropped videos as inputs \cite{Physnet,physformer,physformer++}, (ii) differential normalization method, leveraging frame differences as inputs \cite{DeepPhys,MTTS,metaphys,efficientphys,bigsmall}, and (iii) spatial-temporal map (STMap) method, converting videos to images, with some deep-learning-based methods, utilizing STMap within deep learning networks \cite{APNET}.

Lastly, from a model training viewpoint, the approaches fall into two categories: (i) non-deep neural network (non-DNN) methods \cite{green,ICA,PCA,CHROM,PBV,POS,SSR,LGI,EEMD-MCCA,EEMD+FastICA} and (ii) deep learning methods that utilize various deep learning models.

To address the myriad challenges and issues aforementioned, numerous papers have been published in the rPPG realm. However, a significant fraction of these papers do not openly share their code, thus making a fair evaluation and comparison of the proposed algorithms challenging. The absence of code often compels researchers to reimplement the algorithms from scratch, which leads to potential non-reproducibility due to undisclosed preprocessing and post-processing techniques and specific deep learning model configurations. Moreover, the lack of explicit information regarding train/test/evaluation datasets, such as the time length of the output data used during model evaluation, impedes reproducibility and obstructs fair evaluation and comparisons among different algorithms.

Given these issues, the rPPG research domain tends to be less transparent and more cloistered compared to other research domains. It is highly inefficient for researchers to invest time in reproducing prior works. Hence, this paper aims to provide guidelines for fair evaluation and comparison in rPPG research, and presents an open-source benchmarking framework intended to foster reproducibility and transparency in this field. Through this benchmark project, we aspire to surmount existing limitations, streamline the implementation and evaluation of rPPG algorithms, and enable researchers to have a standardized and comprehensive system for conducting fair and reliable comparisons.

This paper's primary contributions aim to offer objective and comprehensive information and guideline for rPPG research and development, as follows:
\begin{description}
    \item[$\cdot$ rPPG preprocessing techniques:] Given the variance in directory and file structure across public datasets, an efficient interface to access each dataset is essential. Furthermore, the choice of preprocessing methods may vary depending on the deep learning model, such as whether to track faces, crop based on initial face area, crop larger than face size, or crop RoI areas like cheeks and forehead.
    \item[$\cdot$ rPPG dataset lists:] Numerous datasets have been created and made publicly available, but well-organized information about them is lacking. We will provide a comprehensive table detailing each dataset, including supported biometric information, video types (red-green-blue (RGB), near-infrared (NIR)), etc. 
    \item[$\cdot$ rPPG models]: Researchers ofen spend significant time and effort reproducing previous research works. We aim to reduce this time by providing the reference implementation of the key state-of-the-art (SOTA) works in rPPG.
    \item[$\cdot$ rPPG dataset labels:] Typically, datasets provide both photoplethysmography (PPG) and HR labels, but sometimes substantial discrepancies occur between them. Some previous works used HR labels directly from the dataset, while others generated HR labels using methods such as frequency analysis and peak detection.
    \item[$\cdot$ Fair evaluation and comparison:] Some works are not clearly reproduced, and their performance and accuracy cannot be verified. A fairer and more transparent evaluation is achievable by presenting the results obtained through this open-source framework.
\end{description}

\section{Related Works}
In the field of rPPG, there are only a few open-source projects, offering resources for implementing and evaluating rPPG models more efficiently. Such resources enhance the reproducibility and reliability of research findings. Notable examples include:
\paragraph{rPPG-Toolbox\cite{liu2022tool}:} 
The rPPG-Toolbox offers a comprehensice pipeline for rPPG research, development, and evaluation. It incorporates six traditional non-DNN methods, including Green \cite{green}, ICA \cite{ICA}, CHROM \cite{CHROM}, LGI \cite{LGI}, PBV \cite{PBV}, and POS \cite{POS}, alongside six deep learning models, including PhysNet \cite{Physnet} and four subsequent Deepphys models \cite{DeepPhys,MTTS,efficientphys,bigsmall}. The rPPG-Toolbox distinguishes itself by featuring motion augment training and the latest multi-task learning rPPG model, Bigsmall. It also supports preprocessing for various datasets such as SCAMPS \cite{scamps}, UBFC-rPPG \cite{ubfc-rppg}, PURE \cite{pure}, BP4D+ \cite{BP4d}, UBFC-phys \cite{ubfc-phys}, and MMPD \cite{3mmpd}.

\paragraph{iPhys-Toolbox\cite{iphys}:}
Designed for conducting rPPG experiments in the MATLAB environment, iPhys-Toolbox provides a total of six non-DNN methods, including Green \cite{green}, ICA \cite{ICA}, CHROM \cite{CHROM}, POS \cite{POS}, and BCG \cite{bcg} for rPPG analysis and experimentation.

\paragraph{PPG-I Toolbox\cite{ppgi}:}
Launched in 2019 and updated until Oct. 2022, this open-source project focuses on supporting traditional non-DNN methods in rPPG research. Despite the lack of support for deep learning-based methods, it offers capabilities for six traditional non-DNN methods, i.e., Green \cite{green,greenthesis}, spatial subspace rotation (SSR) \cite{SSR}, plane-orthogonal-to-skin (POS) \cite{POS}, local group invariance (LGI) \cite{LGI}, diffusion process (DP) \cite{DP1,DP2}, and Riemannian-PPGI (SPH) \cite{ppgi}, and five evaluation metrics, i.e., correlation, Bland-Altman, root mean square error (RMSE), mean squared error (MSE), and signal-to-noise ratio (SNR). Note that we also consider these metrics in our work because they are useful to assess various aspects of the performance and accuracy of the rPPG algorithms.

\paragraph{pyVHR\cite{pyvhr}:}
It includes its own pipeline for rPPG, offering preprocessing capabilities for 11 open datasets such as PURE \cite{pure}, LGI-PPGI \cite{LGI}, UBFC-rPPG \cite{ubfc-rppg}, UBFC-Phys \cite{ubfc-phys}, ECG-Fitness \cite{hrcnn}, MANOB \cite {MANOHOB}, Vicar-PPG-2 \cite{vicar}, V4V \cite{V4V1st}, VIPL \cite{vipl-hr}, among others. The project provides a variety of methods, including traditional approaches like ICA, PCA, Green, CHROM, POS, SSR, PBV \cite{PBV}, OMIT \cite{omit}, as well as deep learning methods like HR-CNN \cite{hrcnn} and MTTS-CAN \cite{MTTS}. Notably, this project stands out among Python-based open-source projects as it provides an API that offers convenience and ease of use. Additionally, the project provides a visual representation of evaluation metrics, which aids in the analysis and interpretation of results. Overall, this project promotes reproducibility, development, and comparison of rPPG research by providing a comprehensive resource that encompasses various rPPG methods, preprocessing capabilities, open datasets, and evaluation metrics.

\paragraph{PhysBench\cite{physbench}:}
Physbench offers preprocessing capabilities for $7$ rPPG datasets, including RLAP \cite{physbench}, UBFC-rPPG \cite{ubfc-rppg}, UBFC-PHYS \cite{ubfc-phys}, MMPD \cite{3mmpd}, PURE \cite{pure}, COHFACE \cite{cohface}, and SCAMPS \cite{scamps}.
In terms of deep learning models, Physbench provides $5$ different models: DeepPhys \cite{DeepPhys}, TS-CAN \cite{MTTS}, EfficientPhys \cite{efficientphys}, PhysNet \cite{Physnet}, Physformer \cite{physformer}, and Seq-rPPG. Additionally, it offers three non-DNN methods: CHROM \cite{CHROM}, ICA \cite{ICA}, and POS \cite{POS}. It is noteworthy that the Physformer model is implemented using the PyTorch framework, while the other models are developed using TensorFlow. Apart from the provided models and preprocessing capabilities, Physbench also features a dataset generator that synchronizes input videos with corresponding labels. This allows researchers to create synchronized datasets, which can be useful for training and evaluating rPPG algorithms.

\paragraph{Terb\cite{Terbe}:} This project is specifically engineered to run rPPG on the edge device Jetson Nano. It supplies two deep learning methods: DeepPhys and PhysNet. These methods are tailored to run rPPG on the Jetson Nano platform.

\paragraph{bob\cite{bob2017}:} This project offers three non-DNN methods: SSR \cite{SSR}, CHROM \cite{CHROM}, and Li's CVPR14 method \cite{bobref}. These methods are provided to conduct rPPG analysis and experimentation within the project.

\section{Remote Bio-sensing : Open Source Benchmark Framework for Fair Evaluation of rPPG}
\label{sec:headings}

We provide an open-source benchmarking framework that ensures the reproducibility of rPPG algorithms and enables fair evaluation. Unlike the existing frameworks that are limited to certain deep learning and traditional models, our benchmarking framework aims to include all SOTA works - all existing and future research works - in the evaluation list. Furthermore, the framework can be extended for researchers to have the freedom to perform preprocessing, add their own deep learning models, and more comprehensive analysis. Offering such flexibility could be a great contribution, not only to a fair evaluation but also to the advancement of rPPG technology itself.

\begin{table}[ht]
\caption{rppg open-source project list}
    \centering
\begin{tabular}{cccccccccc}
\hline
\multicolumn{2}{c}{Project}                        & \multicolumn{1}{l}{\cite{liu2022tool}} & \multicolumn{1}{l}{\cite{iphys}} & \multicolumn{1}{l}{\cite{ppgi}} & \multicolumn{1}{l}{\cite{pyvhr}} & \multicolumn{1}{l}{\cite{physbench}} & \multicolumn{1}{l}{\cite{Terbe}} & \multicolumn{1}{l}{\cite{bob2017}} & \multicolumn{1}{l}{\textbf{Ours}} \\ 
\hhline{==========}
\multirow{3}{*}{Open Dataset} & Support            & O                                & X                                 & X                                 & O                         & O                             & X                         & X                       & O                        \\ \cline{2-10} 
                              & \# of Dataset      & 6                                & -                                 & -                                 & 11                        & 7                             & -                         & -                       & 6                        \\ \cline{2-10} 
                              & Dataset Analysis   & X                                & -                                 & -                                 & O                         & X                             & -                         & -                       & O                        \\ \hline
non-DNN Method                & \# of Models       & 6                                & 6                                 & 6                                 & 8                         & 3                             & -                         & 3                       & 7                        \\ \hline
\multirow{3}{*}{DNN Method}   & Train Support      & O                                & X                                 & X                                 & X                         & O                             & O                         & X                       & O                        \\ \cline{2-10} 
                              & Evaluation Support & O                                & X                                 & X                                 & O                         & O                             & O                         & X                       & O                        \\ \cline{2-10} 
                              & \# of Method       & 5                                & -                                 & -                                 & 2                         & 5                             & 2                         & -                       & 10                       \\ \hline
\end{tabular}
    \label{tab:tabel1}
\end{table}
Table \ref{tab:tabel1} shows an overview of the characteristics of the previously mentioned open-source projects. Our Open-Source framework stands out by offering a dataset analyzer, which is not available in other open-source projects. Additionally, our framework provides a larger variety of methods compared to other existing tools. This allows researchers to explore and evaluate rPPG algorithms using a wider range of techniques, promoting a more comprehensive analysis and comparison of different approaches.

\begin{figure}[h]
    \centering
    \includegraphics[width=\textwidth]{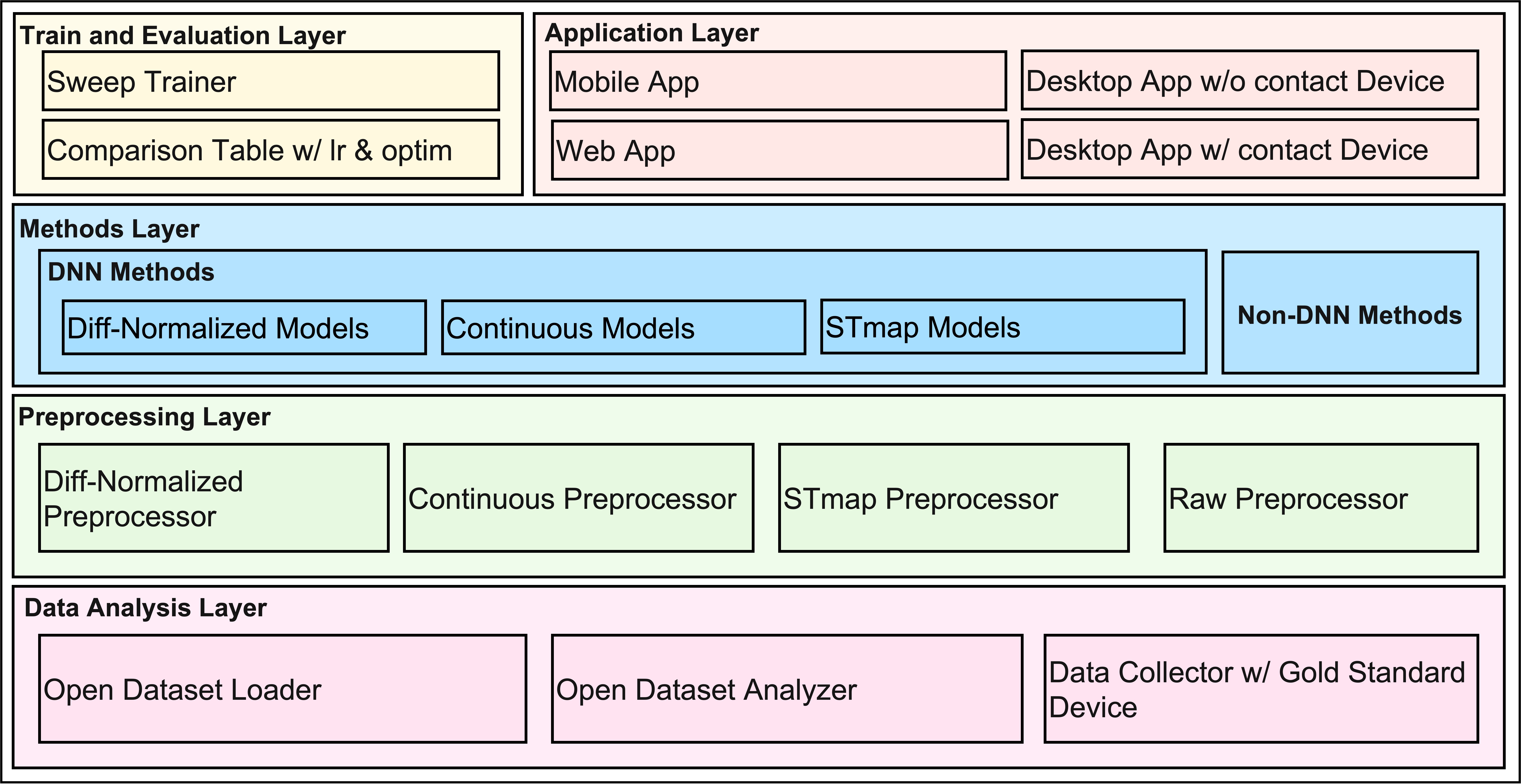}
    \caption{Project Architecture Overview}
    \label{fig:project overview}
\end{figure}

Figure \ref{fig:project overview} provides an overview of the proposed open-source project. It consists of four layers that facilitate the implementation and evaluation of rPPG algorithms.

\begin{itemize}
    \item \textbf{Dataset Analysis Layer:} This layer focuses on dataset analysis and provides tools and functions for analyzing the rPPG datasets. It allows researchers to gain insights into the characteristics and properties of the datasets before conducting further preprocessing and modeling.
    \item \textbf{Preprocessing Layer:} This layer is responsible for preprocessing the input data to make it fit and suitable for the following Model Layer. It includes various preprocessing techniques and functions to handle data normalization, filtering, alignment, and other necessary preprocessing steps.
    \item \textbf{Model Layer:} This layer encompasses the implementation of the main processing of the rPPG technology. It includes both DNN and non-DNN methods. Researchers can choose from a wide range of available models based on their specific requirements and preferences.
    \item \textbf{Train and Evaluation Layer:} This layer facilitates the training and evaluation of the DNN models. It provides tools and functions for training the DNN models using the proper datasets and for evaluating the performance of the model using suitable evaluation metrics.
    \item \textbf{Application Layer:} This layer showcases the application of the implemented model by providing demos and examples. It allows researchers to visualize and interpret the results of the rPPG algorithms in practical applications.
\end{itemize}

Overall, the proposed open-source project offers a comprehensive framework that facilitates dataset analysis, preprocessing, model training and evaluation, and application demonstration, providing researchers with a flexible and integrated platform for their rPPG research and development.

\subsection{Data Analysis Layer}

The Data Analysis Layer provides functionalities related to data collection, retrieval, and analysis. There are over 20 open datasets available, and each dataset has been generated considering different types of information, dimension, artifacts, and noise factors.

\begin{table}[h]
\caption{OpensDataset List}
\centering
\begin{threeparttable}
\begin{tabular}{rclrcc}
\hline
\textbf{\#} & \textbf{year} & \textbf{name} & \textbf{subject} & \textbf{video} & \textbf{label} \\ \hhline{======}
1           & 2011          & MAHNOB\_HCI\cite{MANOHOB}       & 27               & RGB            & ECG\tnote{*}            \\ \hline
2           & 2012          & DEAP\cite{DEAP}                 & 22               & RGB            & PPG            \\ \hline
3           & 2014          & AFRL\cite{AFRL}                 & 25               & RGB            & PPG            \\ \hline
4           & 2014          & PURE\cite{pure}                 & 10               & RGB            & PPG/SPo2\tnote{*}       \\ \hline
5           & 2016          & BP4D+\cite{BP4d}                & 140              & RGB/NIR        & PPG/HR\tnote{*}/BP\tnote{*}      \\ \hline
6           & 2016          & MMSE-HR\cite{MMSE}              & 40               & RGB            & HR/BP          \\ \hline
7           & 2017          & COHFACE\cite{cohface}           & 40               & RGB            & PPG/HR/RR\tnote{*}      \\ \hline
8           & 2017          & BIDMC\cite{BIDMC}               & -                & -              & PPG/BP         \\ \hline
9           & 2018          & LGGI\cite{LGI}                  & 25               & RGB            & -              \\ \hline
10           & 2018          & ECG-Fitness\cite{hrcnn}         & 17               & RGB            & ECG            \\ \hline
11           & 2018          & VIPL-HR\cite{vipl-hr}           & 107              & -              & PPG/HR         \\ \hline
12          & 2018          & OBF\cite{OBF}                   & 100              & RGB/NIR        & PPG/HR         \\ \hline
13          & 2018          & MR-NIRP(ind)\cite{mrnirp_indoor}& 8                & RGB/NIR        & PPG/HR         \\ \hline
14          & 2019          & UBFC-rPPG\cite{ubfc-rppg}       & 42               & RGB            & PPG/HR         \\ \hline
15          & 2020          & VicarPPG\cite{vicar}            & 10               & RGB            & PPG/HR/ECG     \\ \hline
16          & 2020          & MR-NIRP(DRV)\cite{mrnirp_drv}   & 18               & RGB/NIR        & PPG/HR         \\ \hline
17          & 2020          & mori-ppg\cite{morippg}          & 30               & RGB            & ECG            \\ \hline
18          & 2020          & BSIPL-rPPG\cite{pulsegan}       & 37               & RGB            & PPG            \\ \hline
19          & 2020          & EatingSet\cite{vicar}           & 20               & RGB            & PPG/HR         \\ \hline
20          & 2020          & StableSet\cite{vicar}           & 24               & RGB            & HR/HRV/ECG     \\ \hline
21          & 2021          & UBFC-Phys\cite{ubfc-phys}       & 56               & RGB            & PPG/HR/EDA     \\ \hline
22          & 2021          & MPSC-rPPG\cite{mpsc-rPPG}       & 9                & RGB            & HR/RR/HRV      \\ \hline
23          & 2021          & V4V\cite{V4V1st}                & 140              & RGB/NIR        & HR/RR/BP       \\ \hline
24          & 2022          & MTHS\cite{MTHS}                 & 62               & RGB            & PPG/RR         \\ \hline
25          & 2022          & BAMI-rPPG\cite{BAMI}            & 14               & -              & PPG/HR         \\ \hline
26          & 2023          & MMPD\cite{3mmpd}                & 33               & RGB            & PPG            \\ \hline
27          & 2023          & Vital Videos\cite{vitalvideos}  & 900              & -              & -             \\\hline
\end{tabular}
\end{threeparttable}

\label{tab:table2}
\end{table}

Table \ref{tab:table2} presents a list of open datasets. Since the datasets have different structures, it can be time-consuming for new researchers to analyze the data structure. To address this, the Open Dataset Loader assists in easily loading the data based on the analyzed data structure. The Dataset Analyzer offers features such as analyzing the alignment between labels and videos in the open datasets and estimating skin color biases. Additionally, the framework provides features for collecting datasets using a Gold Standard Device in a desktop environment.

\subsubsection{Open Dataset Analyzer}

rPPG utilizes the optical principle of hemoglobin's light absorption to observe vascular changes associated with the contraction and relaxation of blood vessels. However, the degree of variation in reflected light due to light absorption is influenced by the thickness of the skin and the content of melanin. Therefore, it is important to consider these factors as well.

\begin{figure}[h!]
    \centering
    \subfloat[Absorption spectra of melanin in skin and hemoglobin(HbO2) in blood\cite{mathmetical}. Data according to \cite{Laser-tissue}]{\includegraphics[width=0.5\textwidth ]{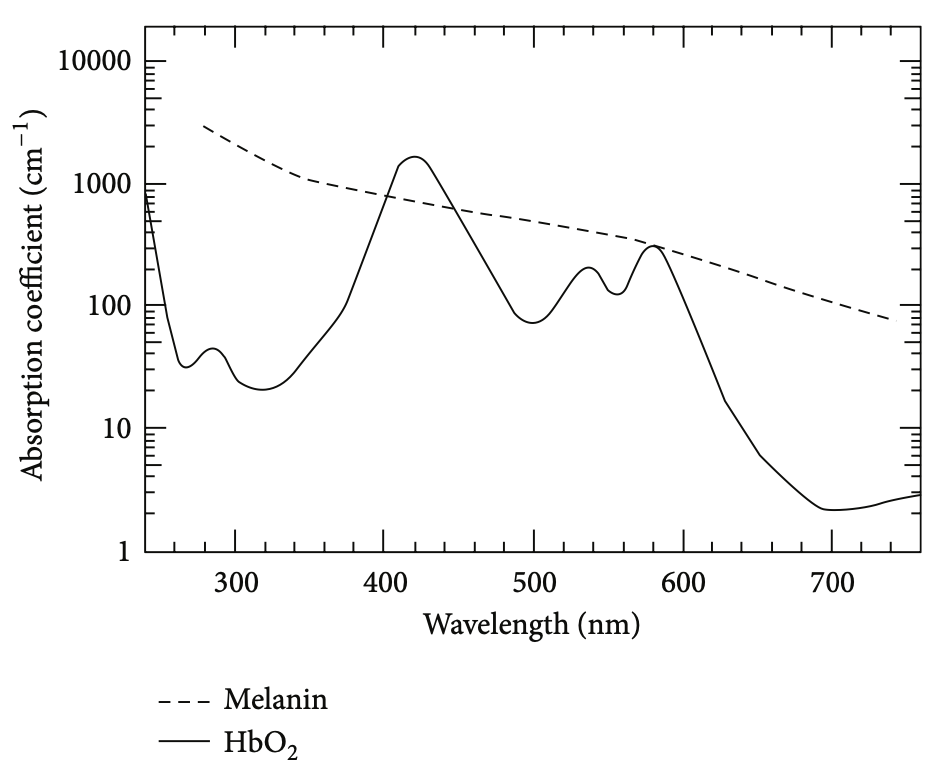}}
    \\
    \subfloat[Camera sensor Spectral Sensitivity\cite{ccd}]{\includegraphics[trim={0 0.1cm 0.5cm 1.6cm} , width=0.5\textwidth, clip ]{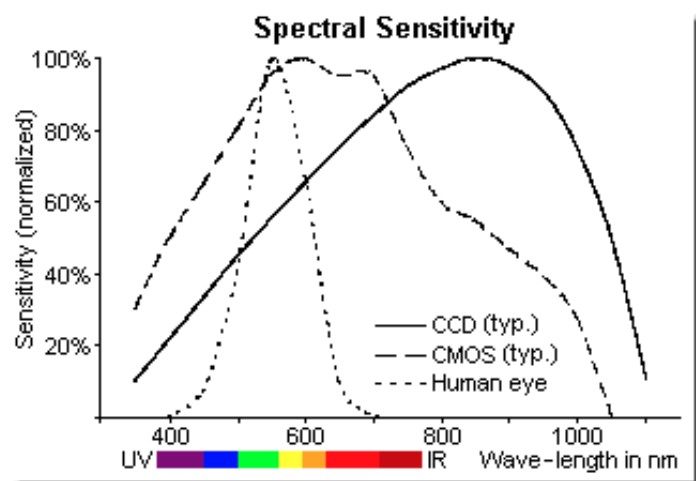}}
    \caption{Graph of light absorption of (a)hemoglobin, melanin, and (b)camera sensor according to the wavelength of light}
    \label{fig:fig2}
\end{figure}

Fig \ref{fig:fig2} represents the absorption spectra of hemoglobin and melanin at different wavelengths. Hemoglobin exhibits the highest light absorption at 432nm, but camera sensors have lower light sensitivity at this wavelength. Therefore, when measuring rPPG, it is necessary to consider wavelengths with both high light sensitivity and high hemoglobin absorption, which typically fall within the range of 500nm to 600nm. However, within this range, melanin has a higher light absorption rate compared to hemoglobin.

\begin{figure}[h]
    \centering
    \includegraphics[width=0.6\textwidth]{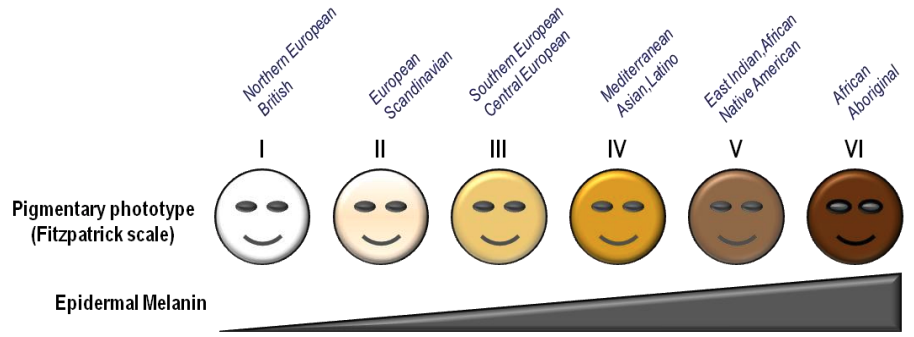}
    \caption{Fitzpatrick scale with Epidermal Melanin \cite{fitz}}
    \label{fig:Fitz}
\end{figure}

Fig \ref{fig:Fitz} illustrates the melanin content based on the Fitzpatrick scale. As the skin type approaches Type 6, the melanin content increases. This can have an impact on the actual measurement results. However, it is rare for open datasets to provide Fitzpatrick scale as a label, making it challenging to ensure fair evaluations. Therefore, to address this issue, it is necessary to measure skintype. However, measuring the Fitzpatrick scale is a difficult task. In our proposed project, we provide a feature that utilizes \cite{fitzeccv} to distinguish skin tones, enabling more accurate analysis.

\begin{figure}[h] 
    \centering
    \includegraphics[width=0.4\textwidth]{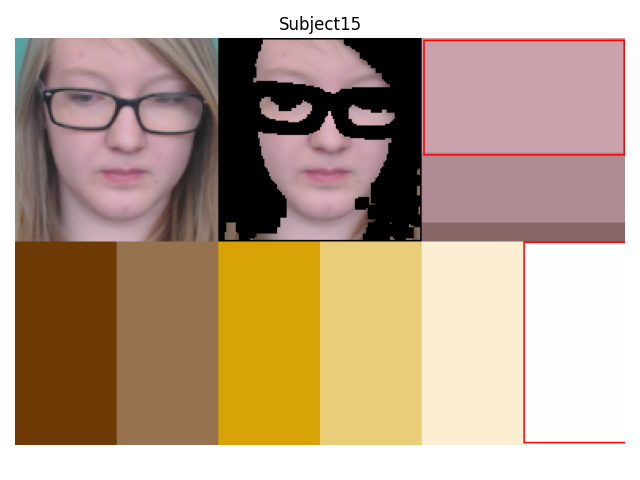}
    \caption{Fitzaptrick scale classification in Open Dataset Analyzer}
    \label{fig:Fitz_15}
\end{figure}

Figure \ref{fig:Fitz_15} depicts the results of performing Fitzpatrick scale classification using our Open Dataset Analyzer. By utilizing the predominant color, we were able to distinguish the most similar Fitzpatrick type among the six types. This classification helps in understanding the skin tone variations within the dataset and provides valuable insights for further analysis and evaluation.

rPPG open datasets typically consist of video data and PPG, HR, RR measurements obtained from a pulse oximeter. The oximeter measures PPG by using wavelengths in the IR and RED bands. It compares the emitted light source with the absorbed values to measure the PPG signal and then processes it through a bandpass filter (BPF) and further post-processing to generate vital signs such as HR and RR.
\begin{figure}[h]
    \centering
    \includegraphics[width=0.8\textwidth]{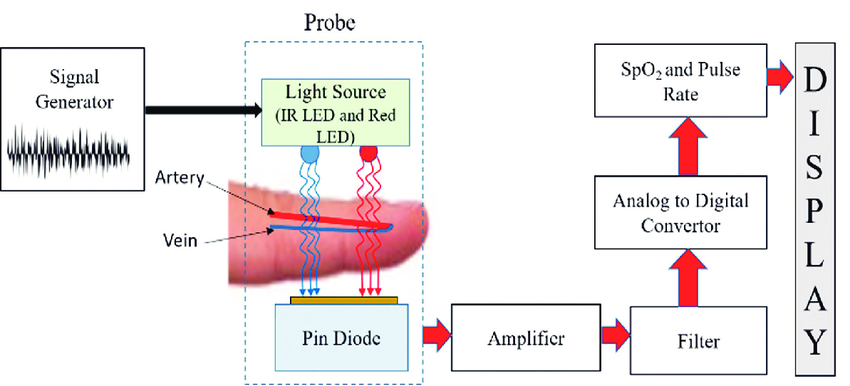}
    \caption{pulse oximeter block diagram \cite{oximeter}}
    \label{fig:oximeter}
\end{figure}

Figure \ref{fig:oximeter} represents a block diagram of a Pulse Oximeter. Generally, when we refer to PPG, it denotes the output of the probe. The internal components of the Oximeter, such as the Filter, ADC, and HR/RR Converter, are responsible for obtaining the HR and respiratory rate (RR) from the PPG signal. However, the settings of the filters and converters (e.g., window size, filter band) can vary across different measurement devices, leading to inconsistencies in the labels provided by open datasets.

To address this issue, when evaluating the performance of algorithms using the PPG labels, it is important to take into consideration the mismatch that can occur between the HR labels and the PPG labels derived from datasets. It is necessary to carefully account for these variations and ensure appropriate alignment between the PPG-based predictions and the corresponding HR labels during evaluation.

\begin{figure}[h]
    \centering
    \includegraphics[width=\textwidth]{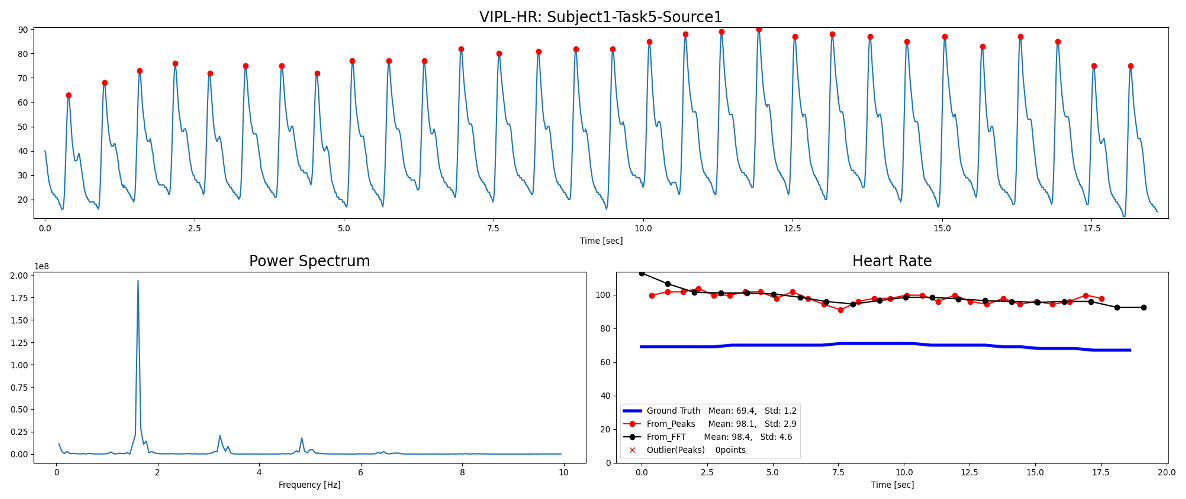}
    \caption{HR estimation at VIPL dataset}
    \label{fig:ppg evaluate}
\end{figure}

Figure \ref{fig:ppg evaluate} shows the results of label evaluation using the Open Dataset Analyzer with the VIPL-HR dataset. Two methods were employed to derive the  HR from the given PPG signal: one using peak detection from the PPG signal and the other using frequency analysis. The bottom right graph illustrates the outcomes. The blue line represents the HR provided as the label, the red line corresponds to the HR derived from peak detection, and the black line represents the HR obtained through frequency analysis. It can be observed that there is some discrepancy between the given HR label and the HR derived from PPG. Some research work uses the given HR labels for ground truth and others rely on the derived HR, which causes confusing results and interpretations.

\subsubsection{Dataset Collector with Golden Standard Device}

TBD

\subsection{Preprocessing Layer}
Preprocessing is a time-consuming and challenging task before training DNN-based rPPG algorithms. With the presence of various open datasets that lack a standardized structure and have different formats, we aimed to simplify the preprocessing process by analyzing the storage formats. Our goal was to enable users to preprocess the datasets without the need to understand the underlying structure, and subsequently save the processed data.

To achieve this, we categorized the preprocessing methods into four approaches:
\begin{itemize}
    \item \textbf{DiffNorm}: DiffNorm preprocesses the data by considering the differences between consecutive frames. It takes advantage of the fact that the reflected light from the skin surface changes due to the pulsatile blood flow. By subtracting the current frame from the previous frame, the variations caused by factors such as lighting conditions, camera noise, and static skin properties can be reduced.
    \begin{equation}
D(t) = \frac{{C'(t) \cdot C(t)}}{{C(t + \Delta t) - C(t)}}
\label{eq:1}
\end{equation}
    Equation represents the preprocessing method of DiffNorm. D(t) denotes the Diff Normalized data at time T, and C(t) denotes the image captured by the camera at time t. It is based on Shafer's dichromatic reflection model and aims to eliminate the quantization noise from the camera and the component of light reflection.
    The goal of DiffNorm is to enhance the pulsatile component of the signal while attenuating the non-pulsatile components. By removing the noise and unwanted reflections, it helps to improve the quality and reliability of the rPPG signal for subsequent analysis and modeling.
    \item \textbf{Z-score normalize}: The Z-score normalize method standardizes the data by subtracting the mean and dividing by the standard deviation. It helps in normalizing the data distribution and reducing the influence of outliers.
    \item \textbf{STmap}: STmap refers to representing the video data as a spatial-temporal map, which captures the spatiotemporal information for further analysis. This representation can help in extracting meaningful features and patterns from the video frames.
    \begin{figure}[h]
        \centering
        \includegraphics[width=\textwidth]{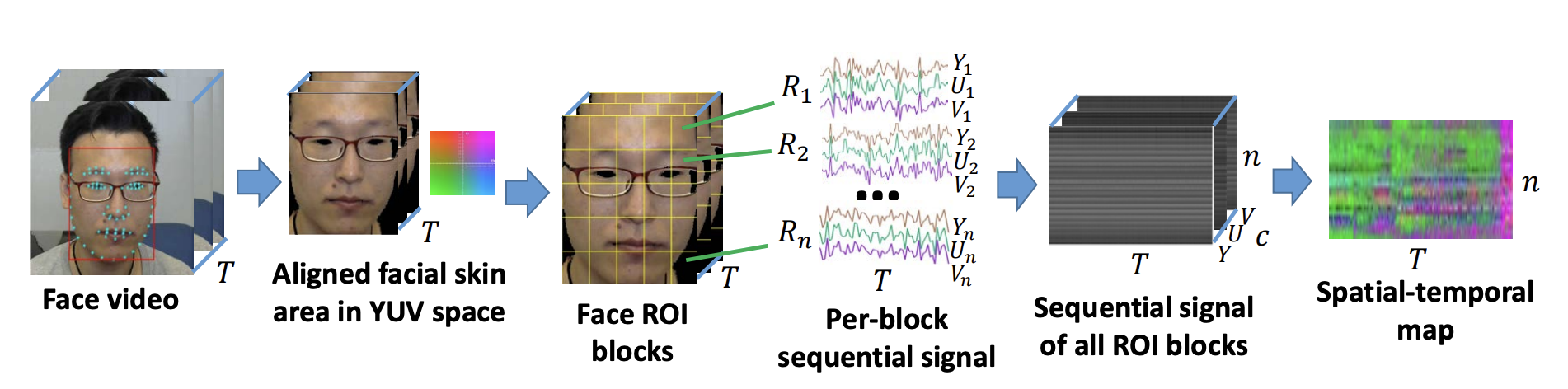}
        \caption{An illustration of spatial-temporal map generation from face video\cite{Rhythmnet}}
        \label{fig:STMap}
    \end{figure}
    Figure \ref{fig:STMap} illustrates the process of generating an STmap from a facial video. STmap, or Spatial Temporal map, incorporates spatial transformation information for each frame of video data into a 2D image. This transformation information describes dynamic changes such as object movement, rotation, and scaling.
    \item \textbf{Raw}: The raw method uses the data in its original form without any preprocessing. This approach is useful when the data is already in a suitable format for further processing.
\end{itemize}

By providing these preprocessing methods, we aimed to alleviate the burden of data preprocessing for users, allowing them to choose and apply the desired preprocessing technique without the need for extensive knowledge of the dataset structure. Additionally, the processed data can be saved for further analysis and training of DNN-based rPPG algorithms.

\subsection{Model Layer}
rPPG is still an active research area using either DNN or non-DNN approaches. Various proposals and efforts are being made in this area. However, some rPPG works lack reproducibility which is not desirable for fair evaluation and further advancement of the technology. 

To address this issue, our goal is to tackle the problem by implementing and openly sharing as many rPPG methods as possible. By providing access to the source code of various methods, we aim to promote transparency, reproducibility, and collaboration within the rPPG research community. We believe that by making the implementations available, we can facilitate knowledge sharing, accelerate research progress, and ultimately contribute to the advancement of the rPPG field.

\subsubsection{Non-DNN Methods}

Table \ref{tab:non-dnn} provides summary of non-DNN rPPG measurement methods which are all non-DNN approaches that leverage the domain characteristics of rPPG to compute the signals. These methods utilize different principles and techniques to extract and estimate the rPPG signal from video data without relying on DNN models.

\begin{table}[h]
\caption{Summary of Non-DNN  rppg measurement methods}
\begin{adjustbox}{width=\textwidth}
\begin{tabular}{c|c|l}
\hline
\textbf{input}                                                      & \textbf{Method} & \multicolumn{1}{c}{\textbf{Representation}}                                                                                                                                                                                                                                                                                                                                              \\ \hhline{===}
\begin{tabular}[c]{@{}c@{}}Raw\\ {[}0 $\cdots$ 255{]}\end{tabular} & GREEN  & \begin{tabular}[c]{@{}l@{}}This method considers the spectral sensitivity of the camera sensor and  the light absorption of hemoglobin. \\ It identifies the Green channel as the most suitable for measuring the specular reflection component. It \\ calculates the average Green channel signal and utilizes predominant frequency analysis to estimate the HR.\end{tabular} \\ \hline
\begin{tabular}[c]{@{}c@{}}Raw\\ {[}0 $\cdots$ 255{]}\end{tabular} & ICA    & \begin{tabular}[c]{@{}l@{}}Independent Component Analysis (ICA) is employed to separate  different signal components mixed within \\ a signal matrix. By applying the JADE algorithm and whitening matrix, the original signals are separated.\\  Empirically, the second separated component is often used as the PPG signal.\end{tabular}                                     \\ \hline
\begin{tabular}[c]{@{}c@{}}Raw\\ {[}0 $\cdots$ 255{]}\end{tabular} & PCA    & \begin{tabular}[c]{@{}l@{}}Principal Component Analysis (PCA) is a well-known dimensionality  reduction technique. In this context,\\  PCA is applied to analyze the main  components of RGB video signals and reconstruct the PPG signal.\end{tabular}                                                                                                                         \\ \hline
\begin{tabular}[c]{@{}c@{}}Raw\\ {[}0 $\cdots$ 255{]}\end{tabular} & CHROM  & \begin{tabular}[c]{@{}l@{}}The CHROM method aims to remove noise caused by light reflection and estimates the PPG signal by \\ filtering out unwanted noise.\end{tabular}                                                                                                                                                                                                       \\ \hline
\begin{tabular}[c]{@{}c@{}}Raw\\ {[}0 $\cdots$ 255{]}\end{tabular} & PBV    & \begin{tabular}[c]{@{}l@{}}PBV proposes a vector that distinguishes motion noise from pulse-related noise in RGB signals \\ to observe variations in heartbeat.\end{tabular}                                                                                                                                                                                                    \\ \hline
\begin{tabular}[c]{@{}c@{}}Raw\\ {[}0 $\cdots$ 255{]}\end{tabular} & POS    & \begin{tabular}[c]{@{}l@{}}POS aims to mitigate specular reflection noise. It projects the PPG waveform onto a plane \\ orthogonal to the skin tone, which helps in signal recovery.\end{tabular}                                                                                                                                                                               \\ \hline
\begin{tabular}[c]{@{}c@{}}Raw\\ {[}0 $\cdots$ 255{]}\end{tabular} & SSR    & \begin{tabular}[c]{@{}l@{}}SSR utilizes the absorbance characteristics of hemoglobin. By applying Subspace Rotation and \\ Temporal Rotation, it extends the pulse amplitude and reduces distortion caused by light reflection.\end{tabular}                                                                                                                                    \\ \hline
\begin{tabular}[c]{@{}c@{}}Raw\\ {[}0 $\cdots$ 255{]}\end{tabular} & LGI    & \begin{tabular}[c]{@{}l@{}}LGI suggests a robust algorithm using differentiable local transformations to handle \\ various environmental conditions.\end{tabular}                                                                                                                                                                                                               \\ \hline
\end{tabular}
\end{adjustbox}
\label{tab:non-dnn}
\end{table}

\subsubsection{DNN Methods}
DNN-based methods in the rPPG field can be categorized based on their input formats, and in our Open-Source Framework, they have been implemented using PyTorch for reproducibility.
\begin{table}[h]
\caption{Summary of DNN  rppg measurement methods}
\begin{adjustbox}{width=\textwidth}
\begin{tabular}{l|c|c|l}
\hline
\multicolumn{1}{c|}{Type}                                                  & input                                                            & Method         & \multicolumn{1}{c}{Representation}                                                                                                                                        \\ \hhline{====}
\multirow{4}{*}{\begin{tabular}[c]{@{}l@{}}Diff\\ Normalized\end{tabular}} & \begin{tabular}[c]{@{}c@{}}Z-score\\ Diff Norm\end{tabular}      & DeepPhys\cite{DeepPhys}       & \begin{tabular}[c]{@{}l@{}}The first deep learning model based on Shafer's Drm. \\ Measure BVP between two $\Delta$ T using two branch CNN\end{tabular}     \\ \cline{2-4} 
                                                                           & \begin{tabular}[c]{@{}c@{}}Avg(Z-score)\\ Diff Norm\end{tabular} & MTTS\cite{MTTS}           & \begin{tabular}[c]{@{}l@{}}Model applying TSM(Temporal Shift Model) to DeepPhys to \\ reduce motion noiseMulti-task model aimed at measuring HR and RR.\end{tabular}      \\ \cline{2-4} 
                                                                           & Diff Norm(Z-score)                                               & EfficentPhys-C\cite{efficientphys} & \begin{tabular}[c]{@{}l@{}}End-to-End model aimed at simplifying the existing DeepPhys \\ and MTTS preprocessing and operating on mobile devices.\end{tabular}           \\ \cline{2-4} 
                                                                           & \begin{tabular}[c]{@{}c@{}}Z-Score\\ DiffNorm\end{tabular}       & BigSmall\cite{bigsmall}       & \begin{tabular}[c]{@{}l@{}}A Wrapping Temporal Shift Model (WTSM) proposal that produces\\ highly accurate results even with a small number of input frames.\end{tabular} \\ \hline
\multirow{3}{*}{Continuos}                                                 & \begin{tabular}[c]{@{}c@{}}Raw\\ {[}-1 .. 1{]}\end{tabular}      & PhysNet\cite{Physnet}        & \begin{tabular}[c]{@{}l@{}}Using negative pearson loss and 3D CNN,\\  rPPG reasoning ability was confirmed.\end{tabular}                                                  \\ \cline{2-4} 
                                                                           & \begin{tabular}[c]{@{}c@{}}Raw\\ {[}-1 .. 1{]}\end{tabular}      & PhysFormer\cite{physformer}     & \begin{tabular}[c]{@{}l@{}}Propose Curriculum Learning Guided Dynamic Loss and \\ verify rPPG inference performance using Transformer.\end{tabular}                       \\ \cline{2-4} 
                                                                           & \begin{tabular}[c]{@{}c@{}}RAW\\ {[}TBD{]}\end{tabular}          & APNET\cite{APNET}          &                                                                                                                                                                           \\ \hline
STmap                                                                      & YUV STmap                                                        & RhythmNet\cite{Rhythmnet}      & Validation of HR Estimation via Video to STmap.                                                                                                                   \\ \hline
\end{tabular}
\end{adjustbox}
\label{tab4:dnnmethods}
\end{table}

Table \ref{tab4:dnnmethods} provides summary of DNN rPPG measurement methods. Deep learning methods are still dependent on the shape of the input, and convergence is determined by the shape of the input.

\subsection{Train and Evaluation Layer}

One of our main objectives is to provide researchers with a convenient research environment. To achieve this, we have incorporated a \emph{fit.yaml} that allows users to easily modify specific configurations, enabling them to customize various aspects of the pipeline, including preprocessing, training, and evaluation. Researchers can make necessary adjustments to suit their experimental setups and research goals by modifying this yaml file. Furthermore, Meta Learning (MAML)\cite{MAML} option can also be configured with proper settings.

In addition, the benchmark's hyperparameters are configured in the \emph{model\_preset.yaml}. This allows for better reproducibility by ensuring that the same hyperparameters can be applied consistently across experiments. Additionally, we have implemented a sweep functionality to facilitate the reproducibility of experiments. Researchers can refer to the Appendix for more detailed information regarding these configurations and implementation details.

To ensure a fair evaluation of the model, we separated the dataset at the subject level, ensuring that the subjects used for training were not used in the evaluation. Additionally, to analyze the model's prediction accuracy over time duration, we conducted evaluations at intervals of [3s, 5s, 10s, 20s, 30s].

The evaluation methods we employed are as follows:

\begin{itemize}
    \item Correlation: Correlation coefficient between the predicted values and ground truth values to assess the linear relationship  
    \begin{equation}
      Correlation = \frac{T\sum_{1}^{T}\hat{y}y - \sum_{1}^{T}\hat{y}\sum_{1}^{T}y }{\sqrt{(T\sum_{1}^{T} \hat{y}^2 - (\sum_{1}^{T} y)^2 ) T\sum_{1}^{T}y^2 - (\sum_{1}^{T}y)^2})}
    \end{equation}
    \item Bland-Altman Analysis: Evaluation of the agreement between the predicted values and ground truth values by examining the mean difference and limits of agreement
    \item Root Mean Square Error (RMSE): A measure of the overall difference between the predicted values and ground truth values
    \begin{equation}
        RMSE = \sqrt{\frac{1}{N} \sum(\hat{y}-y)^2}
    \end{equation}
    \item Mean Absolute Error (MAE): Assessment of the average absolute difference between the predicted values and ground truth values
    \begin{equation}
        MAE = \frac{1}{N} \sum|\hat{y}-y|
    \end{equation}    
    \item Mean Absolute Percentage Error (MAPE) : The average percentage difference between the predicted values and the ground truth values
    \begin{equation}
        MAPE = \frac{1}{N} \sum|\frac{\hat{y} - y}{y}|
    \end{equation}  
    \item Signal-to-Noise Ratio (SNR)\cite{CHROM}: Evaluation of the signal quality of the predicted values compared to the background noise
    \begin{equation}
        SNR = 10log_(10)(\frac{\sum_{f=HR_{min}}^{HR_{max}}(U_t(f)\hat{S}(f))^2}{{\sum_{f=HR_{min}}^{HR_{max}}(1-U_t(f))\hat{S}(f))^2}})
    \end{equation}  where $\hat{S}(f)$ is the spectrum of the pulse signal, S,f is the frequency in beats per minute, and $U_t(f)$ is a binary template window as shown in Fig. 3.
\end{itemize}

By employing these evaluation metrics, we aimed to provide a comprehensive assessment of the model's performance and ensure a fair and objective evaluation of our trained models.

\subsection{Application Layer}
TBD

\section{Reproducibility and Benchmark Results}

In this chapter, we provide some evidence of the reproducibility of our code and present benchmark results for a fair evaluation of each model. For cross-dataset evaluation, we utilized the UBFC and PURE datasets. Deep learning models such as DeepPhys, TSCAN, EfficentPhys, and Bigsmall were evaluated, while non-DNN methods including CHROM and POS were also assessed.

Among the evaluated models, Bigsmall is a multi-task learning method, but we conducted the evaluation using a single-task learning approach. We assessed the performance of each model using evaluation metrics such as MAE (Mean Absolute Error), RMSE (Root Mean Square Error), MAPE (Mean Absolute Percentage Error), and Pearson-Correlation. These metrics were calculated for time periods of 3, 5, 10, 20, and 30 seconds.

    \begin{figure}[h]
        \centering
        \includegraphics[width=0.75\textwidth]{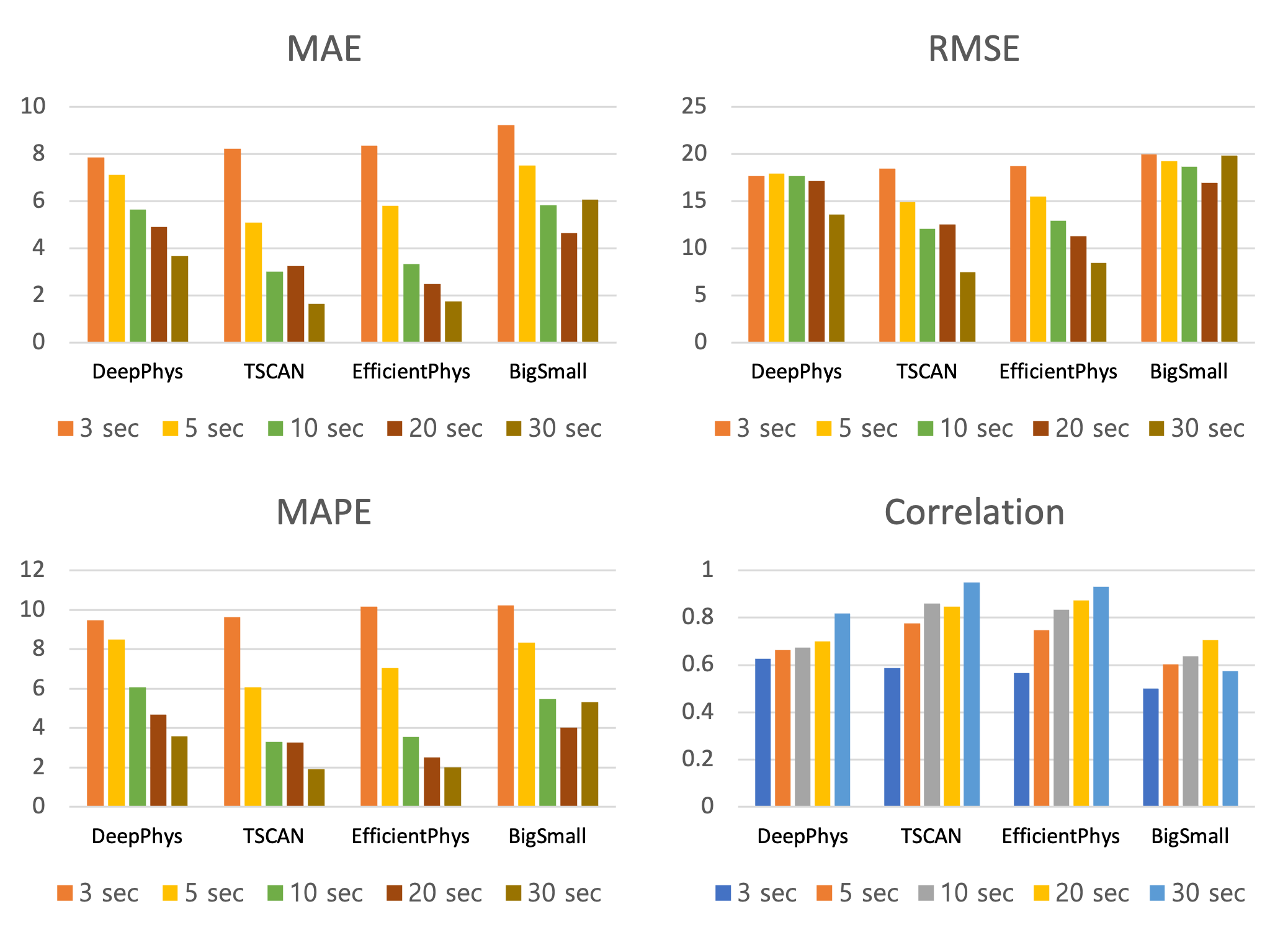}
        \caption{estimate time length evaluation result of UBFC-PURE dataset }
        \label{fig:time_bar_white}
    \end{figure}
Figure \ref{fig:time_bar_white} is the inference result according to the length of time. If the time period is increased from 3 seconds to 10 seconds, the evaluation result improves rapidly, and there is no significant change beyond that, and sometimes gets worse.

    \begin{figure}[h]
        \centering
        \includegraphics[width=0.8\textwidth]{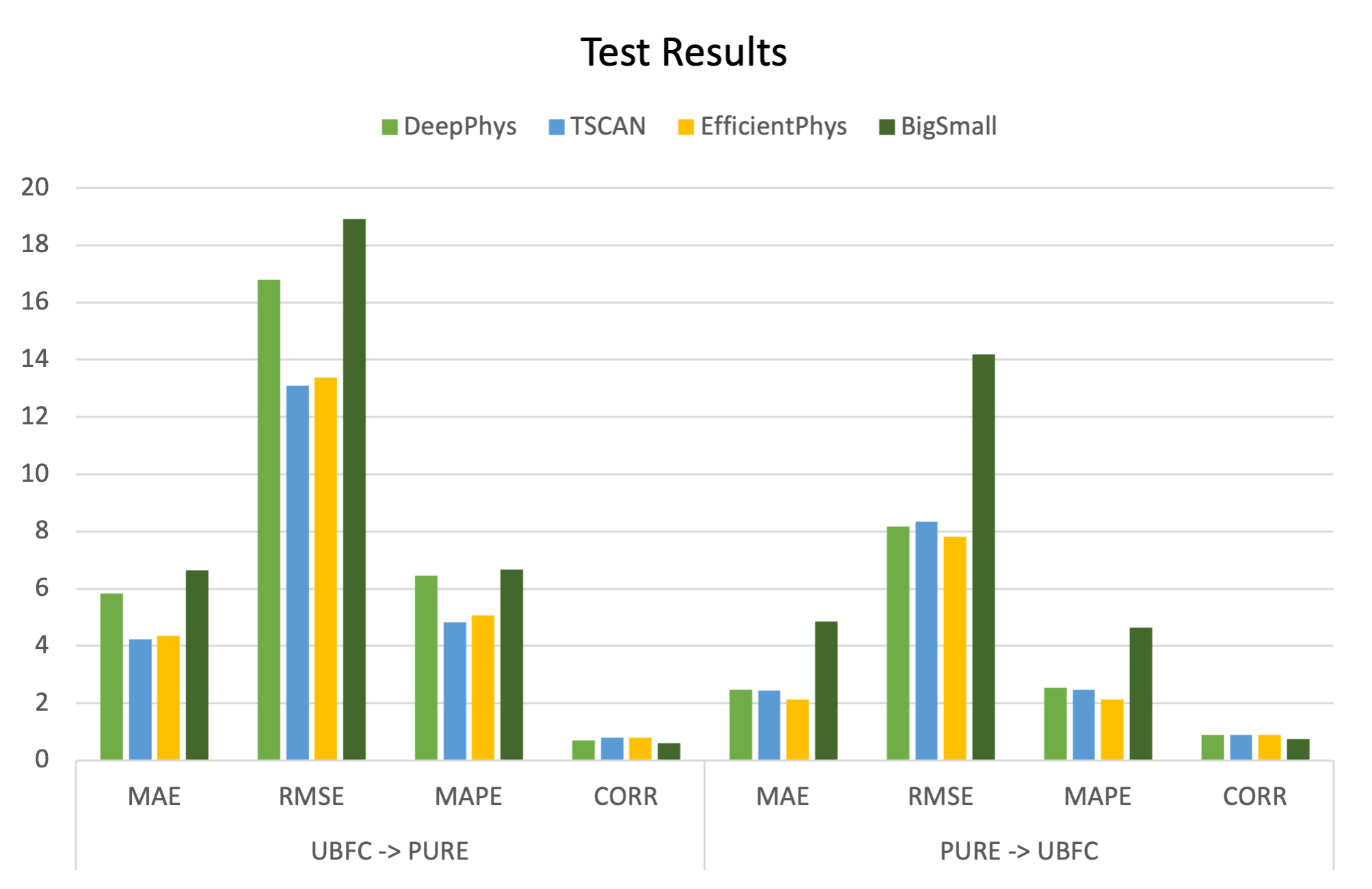}
        \caption{Cross dataset evaluation result of UBFC-PURE, PURE-UBFC dataset (10s)}
        \label{fig:10s}
    \end{figure}

Figure \ref{fig:10s} is the cross-dataset evaluation result. If the dataset has a variety of variables, and the model converges well for the various variables, it can be seen that cross-data set evaluation yields good results. Additional results can be found in the appendix.

With more diverse combinations of building blocks, such as datasets and evaluation metrics, comparison among models will be seen more clearly, and a fair assessment of performance and deeper analysis is possible. Researchers can use these results as a reference for comparing different models and making systematic decisions based on their specific requirements.

\section{Conclusion}

We argue that fair and systematic benchmarking is urgently needed to overcome the challenges in the rPPG technology and make further advancements. To this end, we are proposing a benchmarking framework that enables various rPPG techniques - all existing and future research works - over wide open datasets to be fairly evaluated and compared, including both conventional Non-DNN and DNN methods.

\section{future work}

We intend to keep updating this open-source benchmarking framework that ensures the reproducibility of rPPG algorithms and enables fair evaluation. Important datasets and research works will be continuously added and evaluated accompanied with the comprehensive experiment results and analysis. Furthermore, the framework will be extended for researchers to have the freedom to select preprocessing technology, add their own deep learning models, and conduct a more comprehensive analysis. Offering such flexibility could be a great contribution, not only to a fair evaluation but also to the advancement of rPPG technology itself. We welcome your participation and contribution to this effort.

\section*{Acknowledgments}
This was was supported in part by......

\bibliographystyle{unsrt}  
\bibliography{references}  

\section{Appendix}

\begin{table}[h]
    \centering
    \caption{fit.yaml}
    \begin{adjustbox}{width=0.9\textwidth}
    \begin{tabular}{|l|}
    \hline
    \begin{lstlisting}[style=yaml]
model_save_path: ""     # model save path
preprocess:
  flag: true            # true: preprocess, false: not preprocess
wandb:
  flag: false
  project_name: DeepPhys
  entity: wandb_entitiy

fit:
  model: DeepPhys       # model name
  type: DIFF            # model type ( DIFF, CONT )
  time_length: 180      # model's target length
  overlap_interval: 0   # default 0
  img_size: 72          # model's input size
  train_flag: True      
  eval_flag: True       
  eval_interval: 100    # force evaluation step
  debug_flag: False     # True: debug mode, False: train mode

  train:
    dataset: UBFC
    shuffle: True
    fs: 30              # video fps
    batch_size: 4       # batch size
    learning_rate: 0.009# learning rate
    epochs: 30          # epochs
    loss: MSE           # loss function
    optimizer: AdamW    # optimizer
    meta:               # meta Learning Configuration for MAML
      flag: false
      inner_optim: adam
      inner_loss: MSE
      inner_lr: 0.01

  test:
    dataset: PURE
    shuffle: False
    fs: 30 
    batch_size: 4
    cal_type: FFT       # Heart rate calculation method
    metric: [ 'MAE','RMSE','MAPE','Pearson' ]
    eval_time_length: 10 # second
    \end{lstlisting}
    \\
    \hline
    \end{tabular}
    \end{adjustbox}
\end{table}

\textbf{results}

\begin{longtable}{|c|c|c|c|c|c|c|c|c|}
\hline
\textbf{MODEL} & \textbf{TRAIN} & \textbf{TEST} & \textbf{IMG\_SIZE} & \textbf{Time\_len} & \textbf{MAE} & \textbf{RMSE} & \textbf{MAPE} & \textbf{Pearson} \\ \hline
BigSmall       & PURE           & PURE          & 72                 & 10                 & 0.68         & 1.547         & 0.98          & 0.981            \\ \hline
BigSmall       & PURE           & PURE          & 72                 & 20                 & 0.117        & 0.454         & 0.163         & 0.999            \\ \hline
BigSmall       & PURE           & PURE          & 72                 & 30                 & 0.176        & 0.556         & 0.333         & 0.998            \\ \hline
BigSmall       & PURE           & PURE          & 72                 & 5                  & 1.598        & 3.568         & 2.529         & 0.914            \\ \hline
BigSmall       & PURE           & UBFC          & 72                 & 10                 & 3.419        & 11.862        & 3.338         & 0.817            \\ \hline
BigSmall       & PURE           & UBFC          & 72                 & 20                 & 3.999        & 13.953        & 3.533         & 0.725            \\ \hline
BigSmall       & PURE           & UBFC          & 72                 & 3                  & 6.285        & 15.475        & 6.353         & 0.711            \\ \hline
BigSmall       & PURE           & UBFC          & 72                 & 30                 & 5.323        & 15.329        & 4.851         & 0.69             \\ \hline
BigSmall       & PURE           & UBFC          & 72                 & 5                  & 5.251        & 14.283        & 5.156         & 0.75             \\ \hline
BigSmall       & UBFC           & PURE          & 72                 & 10                 & 5.819        & 18.685        & 5.468         & 0.636            \\ \hline
BigSmall       & UBFC           & PURE          & 72                 & 20                 & 4.634        & 16.923        & 4.015         & 0.706            \\ \hline
BigSmall       & UBFC           & PURE          & 72                 & 3                  & 9.238        & 19.944        & 10.24         & 0.501            \\ \hline
BigSmall       & UBFC           & PURE          & 72                 & 30                 & 6.071        & 19.852        & 5.304         & 0.573            \\ \hline
BigSmall       & UBFC           & PURE          & 72                 & 5                  & 7.516        & 19.226        & 8.346         & 0.603            \\ \hline
BigSmall       & UBFC           & PURE          & 72                 & 10                 & 23.555       & 35.99         & 22.892        & 0.415            \\ \hline
BigSmall       & UBFC           & PURE          & 72                 & 5                  & 23.547       & 35.466        & 24.815        & 0.33             \\ \hline
BigSmall       & UBFC           & UBFC          & 72                 & 10                 & 0.586        & 1.435         & 0.538         & 0.994            \\ \hline
BigSmall       & UBFC           & UBFC          & 72                 & 20                 & 2.539        & 4.184         & 2.43          & 0.947            \\ \hline
BigSmall       & UBFC           & UBFC          & 72                 & 30                 & 0            & 0             & 0             & 1                \\ \hline
BigSmall       & UBFC           & UBFC          & 72                 & 5                  & 0.721        & 2.252         & 0.712         & 0.979            \\ \hline
BigSmall       & UBFC           & PURE          & 72                 & 10                 & 5.718        & 17.785        & 5.532         & 0.677            \\ \hline
BigSmall       & PURE           & UBFC          & 72                 & 10                 & 3.291        & 11.376        & 3.186         & 0.825            \\ \hline
DeepPhys       & PURE           & PURE          & 72                 & 10                 & 0.68         & 1.547         & 1.079         & 0.981            \\ \hline
DeepPhys       & PURE           & PURE          & 72                 & 20                 & 0.117        & 0.454         & 0.163         & 0.999            \\ \hline
DeepPhys       & PURE           & PURE          & 72                 & 30                 & 0.176        & 0.556         & 0.333         & 0.998            \\ \hline
DeepPhys       & PURE           & PURE          & 72                 & 5                  & 1.004        & 2.658         & 1.511         & 0.949            \\ \hline
DeepPhys       & PURE           & UBFC          & 72                 & 10                 & 1.855        & 7.763         & 1.904         & 0.913            \\ \hline
DeepPhys       & PURE           & UBFC          & 72                 & 20                 & 1.516        & 5.287         & 1.557         & 0.957            \\ \hline
DeepPhys       & PURE           & UBFC          & 72                 & 3                  & 4.646        & 12.756        & 4.812         & 0.778            \\ \hline
DeepPhys       & PURE           & UBFC          & 72                 & 30                 & 1.684        & 5.988         & 1.745         & 0.949            \\ \hline
DeepPhys       & PURE           & UBFC          & 72                 & 5                  & 2.609        & 9.021         & 2.647         & 0.884            \\ \hline
DeepPhys       & UBFC           & PURE          & 72                 & 10                 & 5.635        & 17.641        & 6.076         & 0.674            \\ \hline
DeepPhys       & UBFC           & PURE          & 72                 & 20                 & 4.896        & 17.153        & 4.673         & 0.701            \\ \hline
DeepPhys       & UBFC           & PURE          & 72                 & 3                  & 7.857        & 17.698        & 9.472         & 0.627            \\ \hline
DeepPhys       & UBFC           & PURE          & 72                 & 30                 & 3.662        & 13.585        & 3.588         & 0.819            \\ \hline
DeepPhys       & UBFC           & PURE          & 72                 & 5                  & 7.111        & 17.926        & 8.497         & 0.663            \\ \hline
DeepPhys       & UBFC           & PURE          & 72                 & 10                 & 26.719       & 39.369        & 26.05         & 0.178            \\ \hline
DeepPhys       & UBFC           & PURE          & 72                 & 20                 & 25.195       & 39.839        & 22.811        & 0.019            \\ \hline
DeepPhys       & UBFC           & PURE          & 72                 & 5                  & 23.027       & 33.922        & 24.852        & 0.392            \\ \hline
DeepPhys       & UBFC           & UBFC          & 72                 & 10                 & 0.977        & 2.748         & 1.069         & 0.975            \\ \hline
DeepPhys       & UBFC           & UBFC          & 72                 & 20                 & 2.148        & 3.262         & 2.04          & 0.965            \\ \hline
DeepPhys       & UBFC           & UBFC          & 72                 & 30                 & 3.809        & 9.329         & 3.283         & 0.537            \\ \hline
DeepPhys       & UBFC           & UBFC          & 72                 & 5                  & 0.721        & 2.252         & 0.722         & 0.981            \\ \hline
DeepPhys       & UBFC           & UBFC          & 72                 & 10                 & 0.879        & 1.758         & 0.893         & 1                \\ \hline
DeepPhys       & UBFC           & UBFC          & 72                 & 20                 & 0            & 0             & 0             & 1                \\ \hline
DeepPhys       & UBFC           & UBFC          & 72                 & 30                 & 0            & 0             & 0             & 1                \\ \hline
DeepPhys       & UBFC           & UBFC          & 72                 & 5                  & 4.688        & 11.951        & 4.663         & 0.775            \\ \hline
EfficientPhys  & PURE           & PURE          & 72                 & 10                 & 0.567        & 1.412         & 0.94          & 0.991            \\ \hline
EfficientPhys  & PURE           & PURE          & 72                 & 20                 & 0            & 0             & 0             & 1                \\ \hline
EfficientPhys  & PURE           & PURE          & 72                 & 30                 & 0.176        & 0.556         & 0.333         & 0.999            \\ \hline
EfficientPhys  & PURE           & PURE          & 72                 & 5                  & 0.974        & 2.616         & 1.474         & 0.969            \\ \hline
EfficientPhys  & PURE           & UBFC          & 72                 & 10                 & 1.278        & 6.402         & 1.313         & 0.938            \\ \hline
EfficientPhys  & PURE           & UBFC          & 72                 & 20                 & 1.376        & 5.991         & 1.373         & 0.942            \\ \hline
EfficientPhys  & PURE           & UBFC          & 72                 & 3                  & 4.344        & 12.343        & 4.412         & 0.792            \\ \hline
EfficientPhys  & PURE           & UBFC          & 72                 & 30                 & 1.43         & 5.837         & 1.395         & 0.942            \\ \hline
EfficientPhys  & PURE           & UBFC          & 72                 & 5                  & 2.208        & 8.455         & 2.197         & 0.892            \\ \hline
EfficientPhys  & UBFC           & PURE          & 72                 & 10                 & 3.33         & 12.931        & 3.543         & 0.834            \\ \hline
EfficientPhys  & UBFC           & PURE          & 72                 & 20                 & 2.49         & 11.287        & 2.514         & 0.873            \\ \hline
EfficientPhys  & UBFC           & PURE          & 72                 & 3                  & 8.358        & 18.714        & 10.177        & 0.566            \\ \hline
EfficientPhys  & UBFC           & PURE          & 72                 & 30                 & 1.743        & 8.45          & 2.02          & 0.93             \\ \hline
EfficientPhys  & UBFC           & PURE          & 72                 & 5                  & 5.794        & 15.515        & 7.061         & 0.748            \\ \hline
EfficientPhys  & UBFC           & PURE          & 72                 & 10                 & 13.887       & 23.307        & 14.522        & 0.746            \\ \hline
EfficientPhys  & UBFC           & PURE          & 72                 & 20                 & 15.625       & 28.416        & 14.746        & 0.633            \\ \hline
EfficientPhys  & UBFC           & PURE          & 72                 & 5                  & 15.044       & 26.045        & 15.182        & 0.668            \\ \hline
EfficientPhys  & UBFC           & UBFC          & 72                 & 10                 & 0.586        & 2.269         & 0.675         & 0.979            \\ \hline
EfficientPhys  & UBFC           & UBFC          & 72                 & 20                 & 2.197        & 3.479         & 2.035         & 0.95             \\ \hline
EfficientPhys  & UBFC           & UBFC          & 72                 & 30                 & 3.516        & 8.292         & 3.048         & 0.536            \\ \hline
EfficientPhys  & UBFC           & UBFC          & 72                 & 5                  & 0.27         & 1.379         & 0.268         & 0.99             \\ \hline
TSCAN          & PURE           & PURE          & 72                 & 10                 & 0.68         & 1.547         & 1.079         & 0.981            \\ \hline
TSCAN          & PURE           & PURE          & 72                 & 20                 & 0.117        & 0.454         & 0.163         & 0.999            \\ \hline
TSCAN          & PURE           & PURE          & 72                 & 30                 & 0.176        & 0.556         & 0.333         & 0.998            \\ \hline
TSCAN          & PURE           & PURE          & 72                 & 5                  & 0.959        & 2.596         & 1.48          & 0.954            \\ \hline
TSCAN          & PURE           & UBFC          & 72                 & 10                 & 2.296        & 9.068         & 2.315         & 0.884            \\ \hline
TSCAN          & PURE           & UBFC          & 72                 & 20                 & 1.435        & 5.3           & 1.44          & 0.956            \\ \hline
TSCAN          & PURE           & UBFC          & 72                 & 3                  & 4.424        & 12.432        & 4.623         & 0.796            \\ \hline
TSCAN          & PURE           & UBFC          & 72                 & 30                 & 1.634        & 6.089         & 1.488         & 0.942            \\ \hline
TSCAN          & PURE           & UBFC          & 72                 & 5                  & 2.388        & 8.85          & 2.467         & 0.89             \\ \hline
TSCAN          & UBFC           & PURE          & 72                 & 10                 & 3            & 12.098        & 3.286         & 0.859            \\ \hline
TSCAN          & UBFC           & PURE          & 72                 & 20                 & 3.249        & 12.525        & 3.265         & 0.846            \\ \hline
TSCAN          & UBFC           & PURE          & 72                 & 3                  & 8.232        & 18.453        & 9.62          & 0.588            \\ \hline
TSCAN          & UBFC           & PURE          & 72                 & 30                 & 1.628        & 7.435         & 1.924         & 0.948            \\ \hline
TSCAN          & UBFC           & PURE          & 72                 & 5                  & 5.093        & 14.907        & 6.069         & 0.777            \\ \hline
TSCAN          & UBFC           & PURE          & 72                 & 10                 & 24.609       & 37.156        & 24.768        & 0.366            \\ \hline
TSCAN          & UBFC           & PURE          & 72                 & 20                 & 24.805       & 38.792        & 21.923        & 0.417            \\ \hline
TSCAN          & UBFC           & PURE          & 72                 & 5                  & 22.075       & 34.563        & 22.586        & 0.364            \\ \hline
TSCAN          & UBFC           & UBFC          & 72                 & 10                 & 1.367        & 3.612         & 1.48          & 0.955            \\ \hline
TSCAN          & UBFC           & UBFC          & 72                 & 20                 & 2.148        & 3.262         & 2.04          & 0.965            \\ \hline
TSCAN          & UBFC           & UBFC          & 72                 & 30                 & 4.688        & 9.574         & 4.064         & 0.513            \\ \hline
TSCAN          & UBFC           & UBFC          & 72                 & 5                  & 0.361        & 1.592         & 0.368         & 0.989            \\ \hline
TSCAN          & UBFC           & UBFC          & 72                 & 10                 & 0            & 0             & 0             & 1                \\ \hline
TSCAN          & UBFC           & UBFC          & 72                 & 20                 & 0            & 0             & 0             & 1                \\ \hline
TSCAN          & UBFC           & UBFC          & 72                 & 5                  & 4.922        & 13.525        & 4.911         & 0.763            \\ \hline
\end{longtable}

\end{document}